\documentclass[11pt]{article}

\usepackage{amsmath}
\usepackage{amsthm}
\usepackage{epic}
\usepackage{eepic}
\usepackage{epsf}
\usepackage{wrapfig}
\usepackage{latexsym}
\usepackage{program}
\usepackage{times}
\usepackage{witsa4}
\usepackage{multicol}

\newtheorem{example}{Example}[section]

\newcommand{\naive}{na{\"\i}ve}

\newcounter{CommentCtr}

\def\listofcomments{
        \section*{Comments/Corrections\@mkboth{COMMENTS/CORRECTIONS}{COMMENTS/CRRECTIONS}\\{\normalsize Comment number(page number): comment}}
        \begin{description}
        \@starttoc{com}
        \item[\mbox{ }] \mbox{ }
        \end{description}
        \if@restonecol
            \twocolumn
        \fi
}

\newcommand{\PreserveBackslash}[1]{\let\ttemp=\\#1\let\\=\ttemp}

\newcommand{\defeq}{\stackrel{\rm def}{=}}

\newenvironment{myitemize}[0]{%
        \protect\begin{itemize}%
        \vspace*{-2mm}%
        \setlength{\topsep}{-6pt}%
        \setlength{\labelsep}{4pt}%
        \setlength{\partopsep}{-6pt}%
        \setlength{\itemsep}{-2pt}}%
{\protect\end{itemize}}

{\protect\end{enumerate}}

\newcommand{\TEXfigure}[2]{     
        \begin{center}
        \leavevmode
        \input{tex_figs/#1.tex}
        \centerline{\raise 1em\box\graph}
        \caption{\protect #2}
        \label{#1}
        \end{center}
        \end{figure}
}

\newcommand{\qtrue}{{{\rm \bf t}}}
\newcommand{\qfalse}{{{\rm \bf f}}}

{\begin{tabular}{lll}}%
{\end{tabular}}
{\begin{supertabular}{lll}}%
{\end{supertabular}}

\title{Algorithms for Analysing Firewall and Router
  Access Lists}

\author{Scott Hazelhurst\\
Programme for Highly Dependable Systems, Department of Computer Science,\\
University of the Witwatersrand, Johannesburg,Private Bag 3, 2050 Wits,
scott@cs.wits.ac.za}

\date{16 July 1999}

\newcommand{\sa}[2]{\mathit{sa}\mathrm{#1}[#2]}
\newcommand{\da}[2]{\mathit{da}\mathrm{#1}[#2]}
\newcommand{\prt}[1]{\pi_{#1}}

\begin{document}

\pagestyle{empty}

\maketitle

\begin{abstract}\noindent
  Network firewalls and routers use a rule database to decide
  which packets will be allowed from one network onto another. By
  filtering packets the firewalls and routers can improve security and
  performance.  However, as the size of the
  rule list increases, it becomes difficult to maintain and validate
  the rules, and lookup latency may increase significantly.
  Ordered binary decision diagrams (BDDs) -- a compact
  method of representing and manipulating boolean expressions -- are a
  potential method of representing the rules. This paper presents
  a new algorithm for representing such lists as a BDD and then
  shows how the resulting boolean expression can be used to analyse
  rule sets.
\end{abstract}

\section{Introduction}

\label{intro:section}

Network routers and firewalls play a very important role in network
traffic management. By regulating which packets are accepted by a
firewall or router, both the security and performance of the network
can be improved. Routers and firewalls usually have rules which
indicate which packets should be accepted and which rejected. 

A network manager can express many complex rules for accepting or
rejecting packets. The efficacy of traffic management depends on how
good these rules are. In practice, these rules develop over a period
of time and evolve as needs changes. Two problems emerge:
\begin{myitemize}
\item As the list of rules become more complex they become more
  difficult to understand. The person who maintains the list may leave
  and be replaced. For someone to understand the rule base from
  scratch can be very difficult. This makes maintaining the rule base
  difficult if changes are made either for operational or performance
  needs since it may not be obvious what the effect of changing, deleting or
  adding a rule may be. Even changing the position of a rule in
  the rule list can change the semantics of the list.

\item The cost of performing look-up on a rule list may become
  expensive, and particularly for routers, this may add significantly
  to latency in the network.
\end{myitemize}

\paragraph*{Terminology: } To simplify notation, in the rest of the
paper the term \emph{filter} is used to cover both routers and
firewalls, and the term \emph{rule set} is used to refer to a list of
rules that implement the filter's policy.

\subsection{Rule sets}

Filter rules come in several formats; typically these are
proprietary formats. While the expressiveness and syntax of the
formats differ, the following generic description gives a good feeling
for what such rules sets look like. A rule set consists of a
list of rules of the form \(\If |condition| \Then\ |action|\), where
the action is either accept or reject.

\begin{example}\rm
A rule in an access list for a Cisco router might say
something like~\cite{cisco97}:

{\footnotesize
\begin{verbatim}
access-list 101 permit tcp 20.9.17.8  0.0.0.0 
                          121.11.127.20 0.0.0.0 
                           range 23 27
\end{verbatim}
}

\noindent
This says that any TCP protocol packet coming from IP address
20.9.17.8 destined for IP address 121.11.{\linebreak[3]}127.20 is to
be accepted provided the destination port address is in the range
$23\ldots 27$.  More detail is given later.\qed
\end{example}

The rules are searched one by one to see whether the condition matches
the incoming packet: if it does, the packet is accepted or rejected
depending on the action (which will either be accept or reject); if
the condition does not match the rule, the search continues with the
following rules.  If none of the rules match the packet is
rejected.

Since the rules are checked in order, the order in which they are
specified is critical. Changing the order of the rules could result in
some packets that were previously rejected being accepted (and/or
\emph{vice-versa}). 

This paper uses CISCO access-list format for specifying the rule set,
but the methods proposed generalise to other formats.

\subsection{Research goals}

Hazelhurst et al.~\cite{hazel98a} explored the use of binary decision
diagrams (BDDs) for representing access rule lists, and showed the
potential of using BDDs for speeding-up look-up, performing analysis,
and possible hardware support. This paper extends the work focusing
on the problem of analysing the access rule lists.  The contributions
of the paper are:

\begin{itemize}
\item An improved technique for representing a rule set as a boolean
  expression using BDDs;

\item A set of algorithms that can be used to analyse a rule set to help
  validate it, and to understand the effect of changes on the rule set.

\item A prototype graphical interface that can be used to help analyse
  the rule set.
\end{itemize}

\noindent
The use of these techniques will allow more complex and sophisticated
rule sets to be used with greater confidence. This will improve both
security and traffic management.

\subsection{Content and structure of paper}

Section~\ref{bdd:section} introduces the BDDs.
Section~\ref{convert:section} then shows how a rule set can be
converted into a BDD representation. Section~\ref{analyse:section}
explores visualisation and analysis techniques that use BDD
representations.  Section~\ref{graphicsif:section} presents a simple
graphical interface for a prototype filter analysis tool.
Finally, Section~\ref{conclusion:section} describes
the status of the work and discusses possible future research.


\section{Binary Decision Diagrams}

\label{bdd:section}

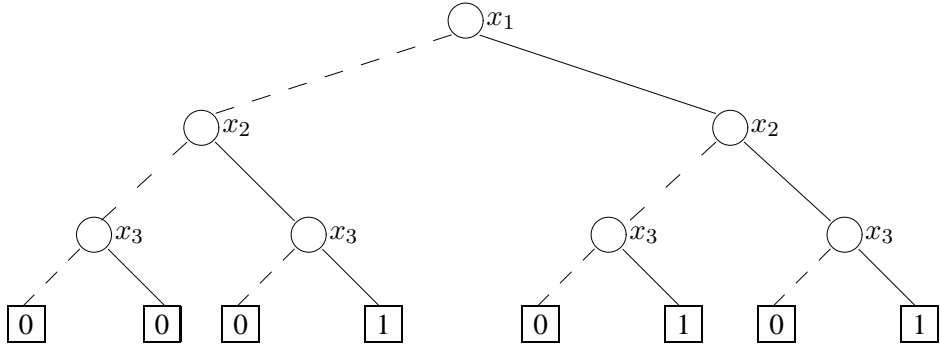
\begin{figure}[ht]
\unitlength=0.95mm
\begin{picture}(160,50)
\put(80,45){\circle{5}\makebox(7,0)[r]{$x_1$}}
\drawline[10](82,43)(115,32)
\dashline[+10]{3}(78,43)(45,32)
\put(43,30){\circle{5}\makebox(7,0)[r]{$x_2$}}
\drawline[10](45,28)(56,17)
\dashline[+10]{3}(41,28)(29,17)
\put(117,30){\circle{5}\makebox(7,0)[r]{$x_2$}}
\drawline[10](119,28)(131,17)
\dashline[+10]{3}(115,28)(103,17)
\put(28,15){\circle{5}\makebox(7,0)[r]{$x_3$}}
\put(16,0){\framebox(5,5)0}
\put(35,0){\framebox(5,5)0}
\dashline[+10]{3}(26,13)(18,5)
\drawline[10](30,13)(38,5)
\put(58,15){\circle{5}\makebox(7,0)[r]{$x_3$}}
\put(46,0){\framebox(5,5)0}
\put(66,0){\framebox(5,5)1}
\dashline[+10]{3}(56,13)(48,5)
\drawline[10](60,13)(68,5)
\put(100,15){\circle{5}\makebox(7,0)[r]{$x_3$}}
\put(88,0){\framebox(5,5)0}
\put(108,0){\framebox(5,5)1}
\dashline[+10]{3}(98,13)(90,5)
\drawline[10](102,13)(110,5)
\put(133,15){\circle{5}\makebox(7,0)[r]{$x_3$}}
\put(121,0){\framebox(5,5)0}
\put(141,0){\framebox(5,5)1}
\dashline[+10]{3}(131,13)(123,5)
\drawline[10](135,13)(143,5)
\end{picture}
\caption{A simple decision diagram for $(x_1\lor x_2)\land x_3$}
\label{bdd:fig:diagram}
\end{figure}

A decision diagram is a method of representing a boolean expression.
For example the boolean expression $(x_1\lor x_2)\land x_3$ can be
represented by the decision diagram in Figure~\ref{bdd:fig:diagram}.

To evaluate an expression given an interpretation of the variables,
you start at the root and move downwards. If the variable has a 0
value, choose the path given by the dashed line; if the variable has a
1 value, choose the path given by the other line. By following this
rule you can easily evaluate the function.

Bryant~\cite{bryant92} introduced the concept of 
reduced, ordered binary decision diagram, which obeys the following rules.

\begin{myitemize}
\item all duplicate terminals are removed (i.e. we shall have at most
  one terminal labelled  1, and one labelled 0);
\item all duplicate non-terminals are removed;
\item all redundant tests are removed (i.e. if both edges leaving a
  vertex go to the same place, you can delete that vertex since it
  implies that the value of the variable that that node represents is
  irrelevant at that point); and
\item a total order is placed on the variables in the expression and
  for any edge (x, y), the label of $x$ comes before the label of $y$
  in the order (variables are encountered in the same order on any
  path from root to leaf).
\end{myitemize}
\noindent
In this paper reduced ordered binary decision diagrams are simply
called binary decision diagrams and abbreviated as BDDs. Efficient
algorithms are known for manipulating boolean expressions (e.g.
conjunction, implication, \ldots). There are two important properties
of BDDs.  First, they are compact representations of boolean
expressions (in a heuristic sense -- there are expressions which are
not compact). Second, for a given variable ordering, the BDD
representation of an expression is canonical. (As a simple example,
this means that if we build BDDs for $\lnot(a\land(b\lor c))$ and
$(\lnot a\lor \lnot b)\land(\lnot a\lor \lnot c)$ we get exactly the
same BDD). In practice this means that checking equivalence can be
done very cheaply once the BDD is constructed. The figure below shows
the BDD representation of $(x_1\lor x_2)\land x_3$.

{
\unitlength=0.9mm
\begin{center}
\begin{picture}(45,40)
\put(30,35){\circle{5}\makebox(7,0)[r]{$x_1$}}
\dashline[+10]{3}(28,33)(22,22)
\drawline[10](32,33)(38,12)
\put(20,20){\circle{5}\makebox(7,0)[r]{$x_2$}}
\dashline[+10]{3}(20,17)(20,5.5)
\drawline[10](22,18)(38,12)
\put(40,10){\circle{5}\makebox(7,0)[r]{$x_3$}}
\dashline[+10]{3}(38,8)(20,5.5)
\drawline[10](40,7.5)(40,5.5)
\put(18,0){\framebox(5,5)}
\put(38,0){\framebox(5,5)}
\end{picture}
\end{center}
}

\noindent 
As can be seen this is significantly smaller. In practice it is not
uncommon to work with expressions that have BDDs tens of megabytes
in size. With such expressions, the efficiency benefits gained by using
BDDs can make many orders of magnitude difference in the size boolean
expressions that can be manipulated.

\subsection*{Number representation}

Integers can be represented as bit vectors, and hence as vectors of
BDDs. Symbolic manipulation of numbers can therefore be done as bit
operations. Among others, addition and comparison can efficiently be
implemented. One of the reasons that many symbolic numerical
expressions can be efficiently implemented as bit-vectors is that BDDs
for common sub-expressions are shared.

\subsection*{Complexity issues}

The size of the BDD is very dependent on the variable ordering
chosen. Although the problem of finding an optimal BDD ordering is
NP-complete \cite{bollig96}, in practice there are good rules of thumb for
finding good variable orderings and many BDD packages come with
heuristic routines for dynamic variable ordering.

It must be emphasised that although BDDs have worked well in many
applications areas, they are not a panacea -- after all the
Boolean Satisfiability problem can easily be represented using BDDs,
which immediately indicates that BDDs cannot be used to solve all
boolean problems efficiently. A stronger result is in fact known ---
there are some problems which require exponential space~\cite{bryant91b}.


\section{Converting rule sets into boolean expressions}

\label{convert:section}

This section describes how a rule set can be converted into a boolean
expression (which is represented as a BDD).
Section~\ref{convert:section:rule} describes how an individual rule in
a rule set can be converted into a boolean expression (and hence a
BDD).  Section~\ref{convert:section:setconv} shows how the boolean
expressions for the individual rules can be combined to give a boolean
expression for the entire list. Some
initial experimental results are given in
Section~\ref{convert:section:results}.

In the description, CISCO access lists are used as
illustration. However the methods can be modified to fit other
approaches.

\subsection{Rule conversion}

\label{convert:section:rule}

A CISCO access rule is of the form

\begin{verbatim}
access-list 101 permit tcp 
            20.9.17.8  0.0.0.0 
            121.11.127.20 0.0.0.0 
            range 23 27
\end{verbatim}

The key components in a rule are:
\begin{myitemize}
\item permit or reject: which indicates packets matching the rule are to be
  accepted. How this field will be used is described in the next
  section.
\item The protocol of the packet: in this case, TCP. Other possible
  examples are UDP and ICMP.
\item The source address: four segments, each a number in the range
  $0 \ldots 255$.
\item The mask for the source address (also four segments).
\item The destination address (in the same format as the source).
\item The destination mask.
\item The range of port addresses. If the port component is empty, all ports
  match. The \texttt{eq x} can be used as short-hand for 
 \texttt{range x x}.
\end{myitemize}

\subsubsection{Representing numbers as bit-vectors}

The key technique used is that numbers can be represented as bit
vectors. For example, an address segment is a number between 0 and
255. At a lower level, the address segment is just a vector of 8
bits. Using BDDs, we can represent sets of numbers symbolically and
perform many operations on them efficiently.

For example, to represent the 8-bit number $x$ symbolically, we
introduce the bit-vector $\langle x_7, \ldots, x_0 \rangle$, where
each of the $x_i$s are boolean (BDD) variables.  The condition that
the vector of $x$'s is equal to 3, is just $\langle x_7, \ldots, x_0
\rangle = \langle
\qfalse,\qfalse,\qfalse,\qfalse,\qfalse,\qfalse,\qtrue,\qtrue\rangle$.
Using the definition of equality of vectors yields the boolean
expression $x'_7x'_6x'_5x'_4x'_3x'_2x_1x_0$, (bits 0 and 1 are high,
the others low). To help make the presentation of formulas more
compact, unless the formula would be confusing, negation is shown
using a prime or tick, and juxtaposition is used for conjunction. The
condition that $\textbf{x}=2$ is represented by
$x'_7x'_6x'_5x'_4x'_3x'_2x_1x'_0$.  The condition
$\textbf{x}=2\lor\textbf{x}=3$ is represented by
$x'_7x'_6x'_5x'_4x'_3x'_2x_1x_0 \lor x'_7x'_6x'_5x'_4x'_3x'_2x_1x'_0$
which is just $x'_7x'_6x'_5x'_4x'_3x'_2x_1$. Large expressions can be
represented compactly using BDDs.

\subsubsection{Boolean variables for the components of a rule}

We introduce a number of boolean variables and expressions to
represent the information in the rule.

We assign each protocol a number $0,\ldots,n_p-1$. These numbers
  can be represented in $m_p=\log_2n_p$ bits, and so we introduce
  $m_p$ variables $\prt{0},\ldots,\prt{m_p-1}$ to encode
  the protocol used. In the examples given below, the protocols can be
  represented in 3 bits.  For example, if the rule refers to a tcp
  protocol (protocol 3) packet, then this is represented by the expression
  $\langle \prt{2},\prt{1},\prt{0}\rangle = \langle
  \qfalse,\qtrue,\qtrue\rangle$, or just
  $\prt{2}'\prt{1}\prt{0}$.

For each segment of the source address we introduce 8 variables
  of the form $\sa{x}{0},\ldots,\sa{x}{7}$,
  where $x$ is the segment number. For example, if segment 2 
  of the source address refers to the number 141, this is
  encoded as 
  $\sa{2}{7}\sa{2'}{6}\sa{2'}{5}\sa{2'}{4}\sa{2}{3}\sa{2}{2}\sa{2'}{1}\sa{2}{0}$.

For each segment of the destination address we introduce 8 variables
  of the form $\da{x}{0},\ldots,\da{x}{7}$,
   where $x$ is the segment
  number. The encoding of destination addresses is similar to the
  encoding of the source address.

As there can be up to 64000 ports specified, port numbers can be
  represented in 16 bits, so we introduce 16 boolean variables
  ($p[15],\ldots,p[0]$, with $p[15]$ being the most significant bit)
  which encode the port number. Using these variables it is possible
  to succinctly represent individual ports as well as ranges of ports.
  Examples are given below.
  
For the moment we ignore the effects of the mask -- the
  Section~\ref{convert:section:masks} discusses mechanisms that
      handle masks.

\subsubsection{Example}

In the example above the source address 20.9.17.8 would be
encoded by the boolean expression:

\hspace*{-3mm}
$\begin{array}{l}
\sa{1'}{7}\sa{1'}{6}\sa{1'}{5}\sa{1}{4}\sa{1'}{3}\sa{1}{2}\sa{1'}{1}\sa{1'}{0}\land\\
\sa{2'}{7}\sa{2'}{6}\sa{2'}{5}\sa{2'}{4}\sa{2}{3}\sa{2'}{2}\sa{2'}{1}\sa{2}{0}\land\\
\sa{3'}{7}\sa{3'}{6}\sa{3'}{5}\sa{3}{4}\sa{3'}{3}\sa{3'}{2}\sa{3'}{1}\sa{3}{0}\land\\
\sa{4'}{7}\sa{4'}{6}\sa{4'}{5}\sa{4'}{4}\sa{4'}{3}\sa{4}{2}\sa{4'}{1}\sa{4'}{0}
\end{array}
$

\noindent
The destination address can be encoded in a similar way.

Representing the range of ports needs a little more care. Let
$\mathbf{port} \defeq \langle p[15],\ldots, p[0]\rangle$. Conditions can
be expressed using boolean operations. Similarly to other
parts of the rule, the condition that the port number must be 25 is
just $\textbf{port}=\emph{int2bv }25$, where \emph{int2bv} is a
function that converts a number into its bit-vector representation.

The range operations can also
be represented efficiently. For example, a greater-than-or-equal-to
operation can easily be defined too. In an ML-like language this might
be defined by:

{\footnotesize
\begin{verbatim}
letrec geq (x:xrest) (y:yrest) =
           (x AND (NOT y)) OR 
           ( (x=y) AND (geq xrest yrest))
   /\  geq [x] [y]  =  (x=y);
\end{verbatim}
}

\noindent
The port-range information in the example rule above would be encoded
as: 

$$(\mathbf{port} \textrm{ geq } \textrm{int2bv}\ 23) 
                    \quad\land\quad
  (\mathbf{port} \textrm{ leq } \textrm{int2bv}\ 27)$$

\noindent
The boolean expression that represents this is
\begin{equation}
\begin{gathered}
(p[15]' p[14]' p[13]' p[12]' p[11]' p[10]' p[9]'
p[8]' p[7]'\\
p[6]' p[5]' p[4])
\land 
(p[3] p[2]'\lor p[3]' p[2] p[1] p[0])
\end{gathered}
\end{equation}

\noindent
This may appear complicated, but the BDD representation is compact.

\subsubsection{Masks}

\label{convert:section:masks}
The source (and destination) addresses in a rule actually both have two
components: a base address and a mask. The mask gives the rule
specifier the flexibility to specify a number of possible matches in
one rule. In effect the mask indicates which bits of the base address
should be matched on and which ignored. Masks are used extensively and
so any mechanism for representing the rule set must be able to deal
with them.

If the base address given in a rule is $s_1.s_2.s_3.s_4$ and the mask is
$m_1.m_2.m_3.m_4$, then a packet with address $a_1.a_2.a_3.a_4$
matches exactly when
\begin{gather}
\label{convert:eqn:mask}
(s_1 \textbf{\,or\,} m_1 \;=\; a_1 \textbf{\,or\,} m_1)\land
(s_2 \textbf{\,or\,} m_2 \;=\; a_2 \textbf{\,or\,} m_2)\land\\
(s_3 \textbf{\,or\,} m_3 \;=\; a_3 \textbf{\,or\,} m_3)\land
(s_4 \textbf{\,or\,} m_4 \;=\; a_4 \textbf{\,or\,} m_4)\notag,
\end{gather}

\noindent
where the \textbf{or} operation is bit-wise or-ing of the two
vectors.\footnote{Here \textbf{or} binds tighter than `='.}
The segments of the mask are typically either 0 (which means the
segment must match exactly) or 255 (which means the segment is
ignored). For example if the source address given is:
\begin{itemize}
\item 146.141.27.66\ \ 0.0.0.0: this means that the packet must match
  exactly as coming from the machine concave.cs.wits.ac.za;
\item 146.141.27.66\ \ 0.0.255.255: this means that the packet must
  come from some machine in the Wits domain.
\end{itemize}

To cope with masks, the mechanism for dealing with addresses described
above needs to be generalised. This is easily accomplished using a
direct implementation of Equation~\ref{convert:eqn:mask}.

\subsection{Conversion of the entire rule set}

\label{convert:section:setconv}

Using the methods described above the entire rule set can in principle
be represented by a boolean expression. Suppose \emph{cvtrule} is
the function that converts one rule into a boolean expression. The
\emph{cvtruleset} function can be defined recursively using
\emph{cvtrule}.

\begin{itemize}
\item If the rule set is empty then no packets can be accepted and so
  the corresponding boolean expression is \qfalse.
    \item If the first rule is an accept rule then a packet will be
      accepted if it matches the rule or if accepted by the rest of the rule
      set. So the corresponding boolean expression is the disjunction
      of the boolean expression representing the first rule and the
      boolean expression representing the rest of the rules.
    \item If the first rule is a reject rule then a packet will be
      accepted if it does not match the first rule and it is accepted by the rest
      of the rule set. So the corresponding boolean expression is the
      conjunction of the negation of the boolean expression
      representing the first rule and the boolean expression
      representing the rest of the rules.
\end{itemize}

\noindent
The pseudo-code for this is given below:

{\footnotesize
\begin{verbatim}
let cvtruleset  ruleset =
      if empty ruleset   return FALSE
      else
        let curr = firstof ruleset 
        let rest = tail    ruleset
        if typeof curr = accept
           return (cvtrule curr) OR 
                  (cvtruleset rest)
        elsif typeof curr = reject
           return (NOT (cvtrule curr)) AND
                  (cvtruleset rest)
\end{verbatim}
}

\subsection{Results}

\label{convert:section:results}

The algorithm described above has been implemented in a protoype tool
built on top of the Voss system~\cite{VossUserGuide}.  This system has
a lazy functional language called FL as its front-end and uses BDDs
internally to represent symbolic boolean expression. The Voss system
also has heuristics for finding good BDD variable orderings.
A simple Perl script processes the rule set which is then read in by
the prototype tool. Then using FL as a front end, a user can analyse
the rules in various ways --- this is described in detail in the next
section. 

The algorithm for converting rule sets into a boolean formula has been
tested one some synthetic test cases and a large real rule set
supplied by an internet service provider.  A set of just over 430
rules provided by a commercial internet service provider was converted
into a boolean expressions using the simple algorithms described
above.  The total time taken to produce the BDD to represent this rule
set was about 20s on a Sun Ultra 4, yielding a BDD of apprimately 1.1K
in size (the text file with the access list is about 32K in size). The
maximum depth of the BDD (determined by the number of variables) is 83
which means that to check whether a packet should be accepted requires
in the worst case 83 bit-operations.

This result is encouraging since it shows that the BDD
representation is feasible and that lookup can be done very quickly.
More experimental evidence is needed, with more rule sets and with
real log data. While worst case is important, average case is much
more important. What average case is depends on what real data looks
like and what the pattern of incoming packets is. This is particularly
important in assessing the cost of the lookup in the
original rule set.

One of the advantages of the boolean expression is that the `shape' of
the BDD has no effect on the semantics, only on the cost of look-up.
So changing the variable ordering may mean that the size of the BDD is
greater or that the length of the maximum path in the BDD grows, but
the semantics remains the same. A statistical analysis of traffic
would indicate a variable ordering to be used that would minimise
average lookup, even if worst-case lookup suffers. Doing the same with
the linear representation of the rules is much more difficult because
of the importance of the linear order of the access lists.



\section{Analysing BDD representations of rule set}

\label{analyse:section}

Section~\ref{convert:section} examined the use of BDDs for compact
representation and lookup in rule sets. This section presents how the
the BDD representation can be used for analysis. While some of the
analysis can be fully automated, the main point of the proposed tool
is to provide a human user with the ability to interact with the rule
set to understand it and the effect of possible changes. The tool does
not act as an oracle, but a means of exploring the rule set.

\subsection{Display of rule set}

The cornerstone of the algorithms to analyse rule sets is the routine
that given a boolean expression representing the rule set displays it
it for a human user. The BDD representation of the rule set is a
compact machine-friendly way of representing the rules; however it is
far from human friendly.

Therefore the tool has an algorithm that presents the boolean formula
in a human readable way, in a tabular form. Here is a simple example
presenting a rule set containing two rules. In the examples that
follow, lines that start with a colon are the input given by the user
to the prototype tool. 

{\footnotesize
\begin{verbatim}
: sc [Proto,Port] cond;
:
Proto Ports      Src 1    Src 2  Src 3   Src 4   Dest1   Dest2  Dest3    Dest4   
  1 | 0--65535 | 0--255  0--255  0--255  0--255 0--255  0--255  0--255  0--255
  3 |       80 | 0--255  0--255  0--255  0--255    120      17     112     100
\end{verbatim}
}

\noindent
The table displays the condition \emph{cond} showing all the values of
source address, destination address, port and protocol which packets
will be allowed through. The first argument to \emph{sc} gives, in
order, the first two columns that should be chosen (the routine has a
default order, but the user can specify any order using the first
argument). The size of the table depends very much of the order of the
columns. Experience has shown that listing using port and protocol
first yields the smallest tables, using the addresses first leads to
huge tables. By changing the order, a user can view the rules in
different ways. At present, displaying the table of a large set of
rules produces very large tables.

\subsection{Instantiating the conditions}

One of the most useful ways of validating a rule set is to ask `what
if' questions. For example:

\begin{itemize}

\item Do we accept packets on port 25? If so what type of packets?

\item On which ports do we accept tcp packets?

\item Which packets do we accept from address $y$?

\item What type of packets will we accept that are being sent to
  address $y$?

\item And so on\ldots
\end{itemize}

\noindent All these queries can be expressed as boolean conditions,
and depending on the user's goal, the results displayed using the
routine described above. Any boolean combination of conditions is
allowed. Here are some examples:

\begin{itemize}

\item What type of udp packets do we accept? 

{\footnotesize
\begin{verbatim}
: sc [Port,Proto] ([Proto <- udp] ::: cond);

Ports Proto  Src 4  Src 3  Src 2  Src 1   Dest4   est3   Dest2   Dest1   
 53 |    2 | 0--255|0--255|0--255|0--255|0--255|0--255 |0--255 | 0--255
\end{verbatim}
}

\item What packets do we accept which have the first segment of the
  destination address of 121 and which are \emph{not} icmp packets?

{\footnotesize
\begin{verbatim}
:  sc [Port,Proto] ([Dest1<-120, NOT(Proto<-icmp)] ::: cond);

Ports Proto Src 4  Src 3  Src 2  Src 1 Dest4  Dest3  Dest2 Dest1   
 0--19|  1|0--255|0--255|0--255|0--255|0--255|0--255|0--255| 120
20--21|  1|0--255|0--255|0--255|0--255|0--255|0--255|0--255| 120
         3|0--255|0--255|0--255|0--255|     3|   112|    17| 120
    22|  1|0--255|0--255|0--255|0--255|0--255|0--255|0--255| 120
         3|     9|     0|    20|   120|0--255|0--255|0--255| 120
23--24|  1|0--255|0--255|0--255|0--255|0--255|0--255|0--255| 120
......
\end{verbatim}
}

\item List the packets we accept for the 120.121 network?

{\footnotesize
\begin{verbatim}
> sc [] ([Source1<-120, Source2<-121] ::: cond);

Ports Proto Src 4  Src 3  Src 2  Src 1 Dest4  Dest3  Dest2 Dest1   
 0--19|  1|0--255|0--255|0--255|0--255|0--255|0--255|0--255| 120
20--21|  1|0--255|0--255|0--255|0--255|0--255|0--255|0--255| 120
         3|0--255|0--255|0--255|0--255|     3|   112|    17| 120
    22|  1|0--255|0--255|0--255|0--255|0--255|0--255|0--255| 120
         3|     9|     0|    20|   120|0--255|0--255|0--255| 120
23--24|  1|0--255|0--255|0--255|0--255|0--255|0--255|0--255| 120
......
\end{verbatim}
}

\item It is also possible using the \emph{gs} routine to summarise the
  results. The \emph{gs} routine lists for each of the columns
  specified in the first argument, the range of values for which
  some packet with the specified condition is accepted. For example:

{\footnotesize
\begin{verbatim}
: gs [Port,Proto,Dest4,Dest3] ([Proto<-tcp] ::: cond);
\end{verbatim}
}

\noindent
shows the range of protocols, protocols and third and fourth
destination address segments for which tcp packets are accepted, without
giving the other details. This is useful as it allows the user to view
the rule set at different levels of abstraction.

\item Which packets are \emph{not} accepted? As the condition for acceptance is
  given as a boolean expression, we can look at its negation to
  discover which packets are not accepted. For example the following
  call to the tool lists, in the order asked for, which gre packets
  destined for ports 80 through 90 are \emph{not} accepted.

{\footnotesize
\begin{verbatim}
sc [Proto,Port,Dest1,Dest2,Dest3,Dest4] 
    ([Port range (80,90), Proto<-gre] ::: not_allowed);
\end{verbatim}
}

\end{itemize}

\subsection{Dealing with modifications}

Probably the most useful part of the tool is the ability to analyse
changes to the rule set, whether those changes are changes to a
particular rule, a change in the order of the rules, or a removal or
addition of a rule. 

The original rule set is represented as a boolean formula, the
modified rule set is represented as a boolean formula, and then the
two formulas together can be used to perform any desired analysis (for
example, whether they two formulas are equivalent, or whether one
logically implies the other). 

An example is given below of the use of the tool. In the example, the
rule set in the file \emph{real1} is read in and stored internally as
a BDD --- it is accessible as the boolean variable \emph{cond}, where
it can be analysed using one of the techniques previously described.
Then the rule set \emph{real1a} is read in and stored internally; at
the same time this is done two boolean variables are set:
\emph{newallow} is a boolean expression which indicates which packets
are allowed by the new rule set but not by the old one, and
\emph{newdeny} indicates which packets are refused by the new rule set
but allowed by the new one. In the example below, those packets are
displayed in a useful order.

{\footnotesize
\begin{verbatim}
: rule ``real1'';
:
: new_rule ``real1a'';
:
: sc [Port,Proto,Dest1,Dest2,Dest3,Dest4,Source1,Source2,Source3] newdeny;

Ports   Proto  Dest1  Dest2  Dest3  Dest4  Src 1  Src 2  Src 3  Src 4
20--21|     3|   120|    17|   21 |     2| 0--119|0--255| 0--255| 0--255
                                              120| 0--11| 0--255| 0--255
                                                      12| 0--207| 0--255
                                                             208|  0--32
                                                                 34--255
                                                        209--255| 0--255
                                                     13|  0--153| 0--255
                                                        155--255| 0--255
                                                     14|  0--226| 0--255
                                                        228--255| 0--255
                                                15--255|  0--255| 0--255
                                       121--255|  0--25|  0--255| 0--255
\end{verbatim}
}

\subsection{Automatic validation}

One automatic validation algorithm has been implemented. This rule goes
through the list of rules and detects any redundant rules --- this can
be done efficiently, and if a redundant rule is detected it is
presented to the user. A redundant rule is not necessarily an error,
but it may result in slower than necessary lookup, and if the user
expects the rule to be useful, then it may indicate that there is a
problem with the rule set.

This detection routine was used on some synthetic examples, and on a
`real' rule set containing approximately 55 rules. In this case, about
5 redundant rules were detected. In most cases the redundant rule is
caused by the same rule appearing more than once in a rule
set. However, another cause of redundancies is caused by mask values
in one rule cover subsequent rules.

Other automatic validation techniques are possible. For example, it
would be possible to show for each (or some) deny rule in a list,
which subequent rules (if any) are affected by it.


\section{User Interface Issues}

\label{graphicsif:section}

The methods presented in the previous sections can be packaged in an
easy-to-use way which hides the underlying algorithm. The first
interface developed was a textual-based interface in FL, which
provides the user with a simple but very expressive query language.

\begin{figure}
\epsfxsize=15cm
\epsffile{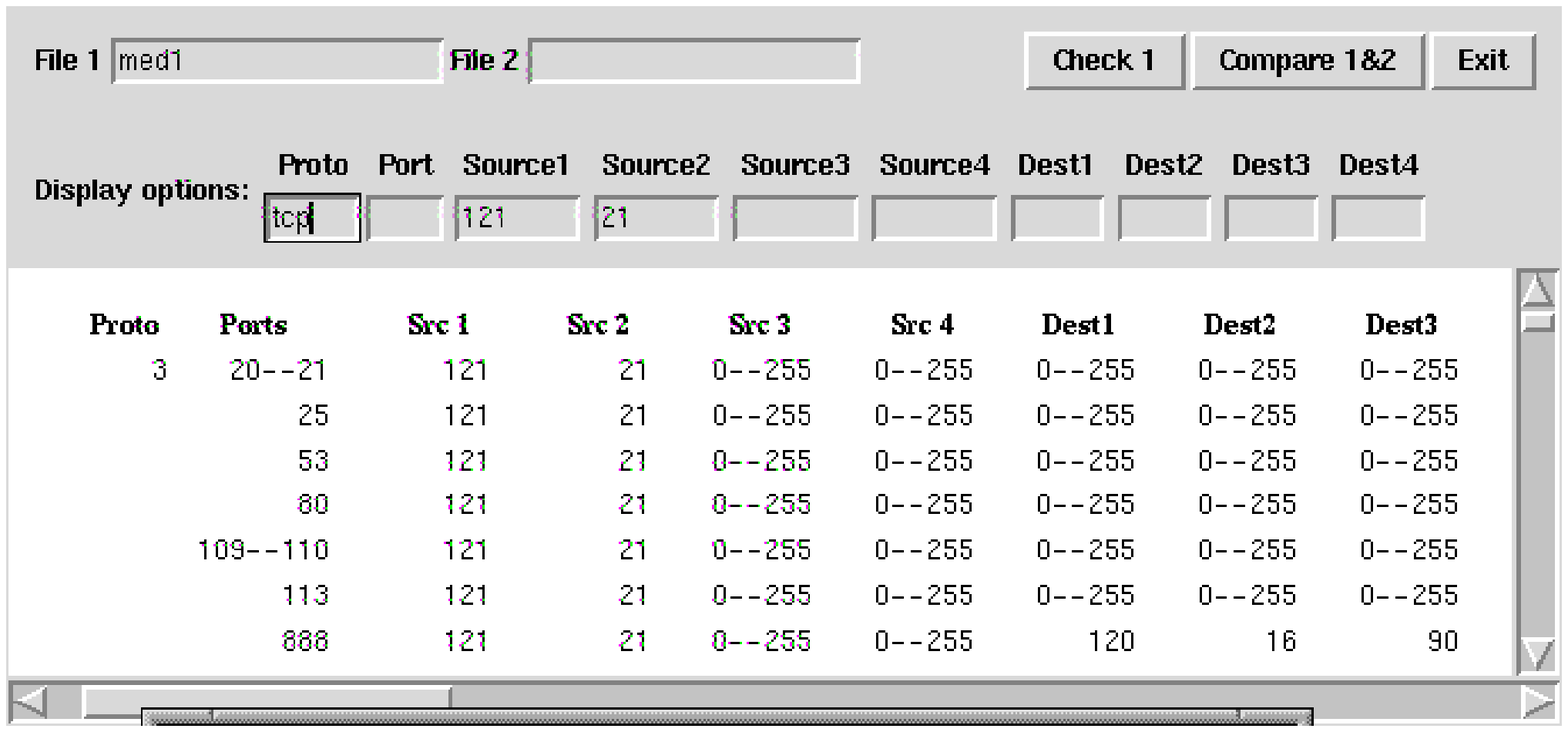}
\caption{Analysing an access file by category}
\label{graphicsif:fig:e1}
\end{figure}

\begin{figure}
\epsfxsize=15cm
\epsffile{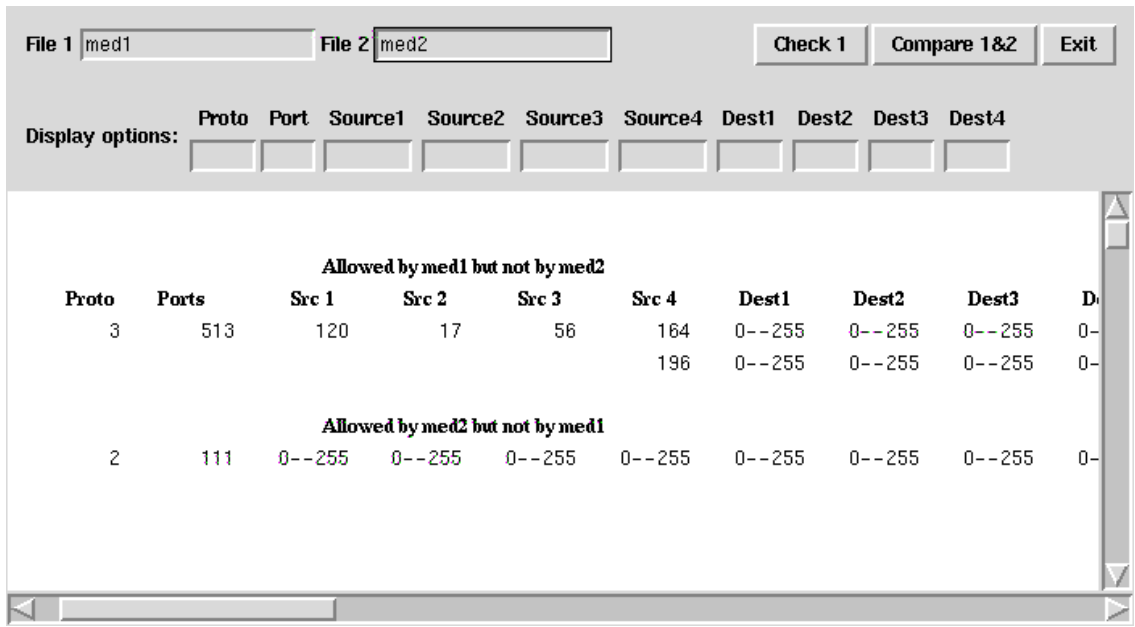}
\caption{Analysing an access file by category}
\label{graphicsif:fig:e2}
\end{figure}

A prototype graphical interface has been developed using Tcl/Tk and is
illustrated in Figures~\ref{graphicsif:fig:e1} and
\ref{graphicsif:fig:e2}.  Figure~\ref{graphicsif:fig:e1} shows the
analysis of an access file. The user types in the name of the file in
the given box. Below that is information showing how the analysis will
be shown. `Display options' shows the order in which the information
will be presented. By clicking on these options the user can change
the order of presentation.  The user can also restrict the analysis by
entering values in the boxes. In the example given, only rules
pertaining to tcp packets from the 121.21 network are displayed.

Figure~\ref{graphicsif:fig:e2} shows the results of comparing two rule
sets. The user enters the file names and chooses the appropriate options
for displaying the results. The differences are then displayed in
the window.

For an industrial strength front-end, more care needs to be taken with
the design. The ideal system would probably provide the most important
functionality through the graphical interface, and then allow an
advanced user to issue more powerful queries through a textual interface.


\section{Conclusion}

\label{conclusion:section}

This paper has examined the problem of using filter rule sets for
routers and firewalls. These rule sets are important for both security
and performance. Unfortunately as the size of the filter rule sets grow it
becomes more difficult to understand the rules.

This paper has shown that even large rule sets can be represented as a
boolean expression in a compact way using BDDs. As a boolean
expression it can easily be manipulated in various ways which allows
the rule set to be manipulated an analysed. 

\begin{itemize}
\item The rule set can be displayed at various levels of abstraction
  from different perspectives. This enables the user to understand
  what the rule set allows and does not allow.

\item A range of queries can be performed on the rule set. This allows
  a human user to test the rule set to ensure that the behaviour of
  the rule set is as expected.

\item The effect of changing the rule set can easily be seen. This can
  help reduce the possibility of errors being introduced.

\item Some automatic analysis of the rule set is possible.

\item A simple graphical interface to the tool enables the
    algorithms presented in this paper to be used easily without the
    user having to understand the underlying theory.
      
\end{itemize}

In all cases, the computational resources required are modest.
By using these techniques a network manager can gain greater
confidence that the rule set is correct. This will also allow larger
and more complex rule sets to be used, improving both the performance
and the security of the network.

There are a number of areas for future research. How to present the
rules in tabular form in a compact way needs further work. A {\naive}
algorithm works reasonably well, but the table size can grow
dramatically. It should be possible to present the table more
compactly. Also by integrating the tool with other tools like nslookup
it should be possible present the information in a friendlier way.
(From a practical point of view, a lot of work could be spent on the
graphical interface.) Another possibility is using the boolean formula
to generate the set of rules as a CISCO access-list (using only accept
or only deny rules). 

The prototype tool is not efficiently implemented
(it is a collection of C, FL, Perl and Tcl/Tk code). While acceptable
for a prototype, an industrial-strength tool would  need to be
efficiently reimplemented.

More experience and case studies could also lead to other ways of
automatically analysing rule sets.

Finally, there are related research areas: how to use the BDD
representation for fast lookup, and possible hardware support.

\subsection*{Acknowledgements}
I gratefully acknowledge the help of The Internet Solution who posed
this question initially and who provided examples of real access lists
to us. Andrew Henwood and Anton Fatti worked on a previous version
of this tool. The work was supported by grants from the University of
the Witwatersrand Research Committee and the National Research
Foundation. Some of the work was done while the author was on
sabbatical leave at the Laboratoire D'Analyse et D'Architecture des
Syst\`emes, Centre National de la Recherche Scientifique.
A shortened version of the paper appears in the South African
Telecommunications and Networks Application Conference '99.


\bibliography{refs}

\begin{thebibliography}{1}

\bibitem{bollig96}
B~Bollig and I~Wegener.
\newblock `{Improving the Variable Ordering of OBDDs is NP-Complete}'.
\newblock {\em {IEEE Transactions on Computers}}, {\bf 45}(9):993--1002,
  (September 1996).

\bibitem{bryant91b}
R~Bryant.
\newblock `{On the Complexity of VLSI Implementations and Graph Representations
  of Boolean Functions with Application to Integer Multiplication}'.
\newblock {\em IEEE Transactions on Computers}, {\bf 40}(2):205--213, (February
  1991).

\bibitem{bryant92}
R~Bryant.
\newblock `{Symbolic} {Boolean} {Manipulation} {with Ordered} {Binary-Decision}
  {Diagrams}'.
\newblock {\em ACM Computing Surveys}, {\bf 24}(3):293--318, (September 1992).

\bibitem{cisco97}
{Cisco Systems Inc.}
\newblock {Configuring IP Systems}.
\newblock Published at the Cisco web site, 1997.
\newblock http:// www.cisco.com /univercd /cc /td /doc /product /software.

\bibitem{hazel98a}
S~Hazelhurst, A~Fatti, and A~Henwood.
\newblock `{Binary Decision Diagram Representations of Firewall and Router
  Access Lists}'.
\newblock Technical Report TR-Wits-CS-1998-3, {Department of Computer Science,
  University of the Witwatersrand}, (October 1998).
\newblock Proceedings of SAICSIT '98.

\bibitem{VossUserGuide}
C~J Seger.
\newblock `{Voss --- A Formal Hardware Verification System User's Guide}'.
\newblock Technical Report 93-45, Department of Computer Science, University of
  British Columbia, (November 1993).
\newblock Available by anonymous ftp as
  ftp://ftp.cs.ubc.ca/pub/local/techreports/1993/TR-93-45.ps.gz.

\end{thebibliography}

\end{document}